\documentclass{PoS}

\title{Radio signals of particle dark matter}

\ShortTitle{Radio signals of particle dark matter}

\author{\speaker{Marco REGIS}
       University of Turin - INFN\\
       E-mail: \email{regis.mrc@gmail.com}}

\abstract{In most of particle dark matter (DM) models, the DM candidate injects sizable fluxes of high-energy electrons and positrons through its annihilations or decays. Emitted in regions with magnetic field, they in turn give raise to a synchrotron radiation, which typically covers radio and infrared bands. We discuss the possibility of detecting signatures of Galactic and extra-galactic DM in the total intensity and small-scale anisotropies of the radio background.}

\FullConference{XXIst International Europhysics Conference on High Energy Physics\\
                21-27 July 2011\\
                Grenoble, Rhones Alpes France}

\begin{document}

\section{Introduction}
From the title of my talk, one may wonder how it can be possible to infer something completely unknown, as non-gravitational signals of dark matter (DM), 
by means of something quite well-known, as radio-astronomy.
Discovery of new physics normally goes through designing new experiments. Here the idea is very simple, namely, that the technique of radio observations is potentially very promising in the quest for particle DM, but requires improved flux and angular sensitivities with respect to current capabilities.
The development of ASKAP, EVLA, MeerKAT, and, in particular, SKA makes the near future promisingly bright in this respect.
Those new radio telescopes will certainly discover signals of previously unknown physical mechanisms, and we discuss the possibility that an emission induced by particle DM will be among them.

Weakly interacting massive particles (WIMPs) are the most investigated class of DM candidates in the literature (for a review, see, e.g.,~\cite{Bertone:2010zza}).
One of the routes to test the hypothesis of WIMP DM stems from the 
the bases of the framework themselves. Indeed, given the weak interaction, there is a ("weak" but finite)
probability that WIMPs in DM halos annihilate in pairs or decay into detectable species. 
In particular, and with the exception of WIMP models annihilating/decaying into neutrinos only, a sizable branching ratio of annihilation/decay into electrons and positrons is a general feature of WIMP models (see, e.g., Fig.~4 in \cite{Regis:GC}).
Interactions of high-energy $e^+/e^-$ with the interstellar
magnetic field in astrophysical structures give rise to magnetic bremsstrahlung called synchrotron radiation.
In the monochromatic approximation of synchrotron power, the energy of an electron emitting at frequency $\nu$ is given by $E\simeq 15\sqrt{\nu_{GHz}/B_{\mu G}}$ GeV (where $\nu_{GHz}$ is the frequency in GHz and $B_{\mu G}$ is the magnetic field in $\mu G$). It follows that, assuming a magnetic field of few $\mu G$ (as typical for galaxies), emissions at radio frequencies are mostly generated by electrons with energy around 1-10 GeV.
Therefore, assuming a DM mass $\gtrsim 10$ GeV, a sizable DM-induced radio synchrotron emission is a general prediction of WIMP models (with, of course, spectrum and absolute flux depending on the specific model).

\section{Targets}
We now investigate possible targets for detection of DM radio emission and see that in some cases the benchmark `thermal' annihilation rate $ (\sigma_a v) =3\cdot 10^{-26}{\rm cm^3s^{-1}}$ can be already probed.

{\bf Galactic Center: } Although the nucleus of our Galaxy is a very rich system, where a clean disentanglement of a DM signal from astrophysical emissions is rather complicated, the Galactic Center (GC) is one of the prime targets in WIMP indirect searches, given the large DM overdensity predicted by N-body numerical simulations. Observations of Sgr A$^*$ are not consistent with a DM interpretation and the GC inner-part allows to constrain DM masses below few tens of GeV for a thermal annihilation cross section~\cite{Regis:GC,GC}. Slightly larger scales ($\sim1^\circ$) could set stronger constraints~\cite{Regis:GC} but dedicated observations would be in order. On the scale of galactic bulge, a hint for a DM signal has been suggested~\cite{Haze} (the so called ``WMAP Haze''), which however needs further investigation, given sizable uncertainties related to relevant astrophysical components.

{\bf Galactic Halo:} Mid-high latitudes can be a cleaner test since they involve propagation of $e^+/e^-$ far from the GC and galactic disc, and so magnetic field and transport parameters suffer of smaller uncertainties. Searches have to be focused on low radio frequencies in the case of WIMP models inducing an $e^+/e^-$ spectrum softer than the galactic cosmic-ray one, while at microwave in the opposite scenario.
No evidences, but interesting constraints have been derived for PAMELA DM candidates (i.e., heavy and leptophilic WIMPs)~\cite{Halo}.
Recently, it has been also shown that for thermal annihilation rate and cuspy profiles, low-frequency galactic radio emission constrains DM candidates with $M_{DM}\lesssim10$ GeV~\cite{Fornengo:2011iq}.

\begin{figure}[t]
\includegraphics[width=0.45\textwidth]{ARCADE_fits_1.eps}
\hspace{0.8cm}
\includegraphics[width=0.45\textwidth]{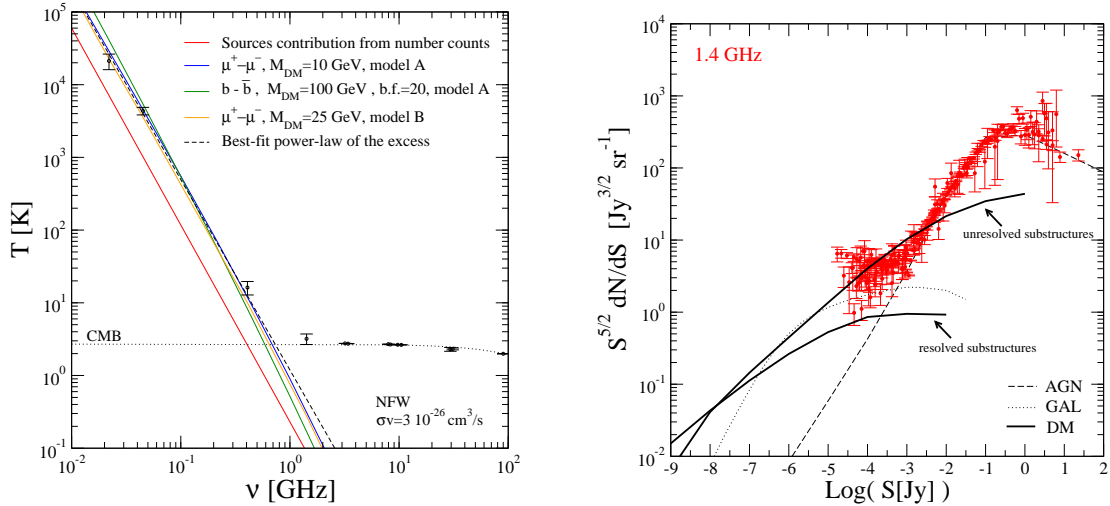}
\caption{{\it Left:} Extragalactic radio background as derived by ARCADE~\cite{Fixsen:2009xn}, together with 
three possible interpretations of the low--frequency ($<10$ GHz) excess in terms of WIMP annihilations (blue, green, and orange curves, see Ref.~\cite{Fornengo:2011cn} for details). 
The astrophysical source contribution estimated from number counts (red line), the CMB contribution (black--dotted line), and a best--fit power--law of the excess (black--dashed line) are also reported. 
{\it Right:} Differential number counts for AGNs (dashed line), star--forming galaxies (dotted line), and the same 10 GeV benchmark DM model (solid lines) shown in left panel. For DM, we consider a case such that all substructures are resolved, and an opposite case where all substructures are unresolved.
Plots are from Ref.~\cite{Fornengo:2011cn}.
}
\label{fig:arcade}
\end{figure}

{\bf Extragalactic diffuse emission:}
The ARCADE 2 Collaboration has recently measured an isotropic radio emission which is significantly brighter than the expected contributions from known extra--galactic sources~\cite{Fixsen:2009xn}. The simplest explanation of such excess involves a ``new'' population of unresolved sources which become the most numerous at very low (observationally unreached) brightness. In particular, the scenario which best fits is a population of numerous and faint synchrotron sources generated by primary electrons with a hard spectrum and with no or very faint correlated mechanisms at infrared and gamma--ray frequencies.
In our current understanding of structure clustering, any luminous source is embedded in a DM halo, and therefore extragalactic DM halos can be seen as the most numerous source population. The flux induced by WIMP annihilations/decays is predicted to be very faint. It is associated to primary electrons and positrons generated as final state of annihilation/decay, and WIMP models with large annihilation/decay branching ratios into leptons induce hard spectra of $e^+/e^-$ with very faint gamma--ray counterpart (and, of course, no straightforward thermal emission).
In Ref.~\cite{Fornengo:2011cn}, the observed excess is investigated in terms of synchrotron radiation induced by WIMP annihilations or decays in extragalactic halos, concluding that, intriguingly, for light--mass WIMPs with thermal annihilation cross-section, the level of expected radio emission matches the ARCADE observations, as shown in Fig.~\ref{fig:arcade}.
Further tests with next generation radio-telescopes on extragalactic number counts (see Fig.~\ref{fig:arcade}) and small-scale anisotropies~\cite{Zhang:2008rs} can provide new informations for addressing such hint.

{\bf Extragalactic objects:}
A non-thermal emission with a spherical morphology correlated with the DM halo profile inferred from kinematic measurements in the external part of extragalactic objects can be a strong evidence for WIMP induced emission. Most promising targets are dwarf spheroidal galaxies and clusters of galaxies (see, e.g.,~\cite{Colafrancesco}). Dwarf spheroidal galaxies are an ideal target since they are the most DM dominated objects discovered in the local Universe, and they are also the faintest and most metal-poor stellar systems known (which implies a very low expected background from the baryonic component). The first dedicated radio campaign has been recently completed~\cite{dwarf} with the observation of six nearby dwarf spheroidal galaxies with the ATCA telescope array~\cite{atca}. Data analysis is in progress.

\section{Conlcusion}

In order to isolate a non-gravitational DM contribution from astrophysical background, large sensitivities and angular resolutions are essential. In particular, clear evidences may arise by observing the outer part of galaxies (along the same line pursued for gravitational search of DM in rotation curves).
For this purpose radio astronomy is a superior tool among indirect DM searches through non-thermal diffuse-emission channels.
Therefore, although probably radio signals won't be the last word on the DM nature (given large uncertainties associated to the $e^+-e^-$ transport and magnetic fields, which mine the possibility of tracing back to DM microscopic properties), they could be one of the first evidences of non-gravitational DM interactions.

\acknowledgments

MR acknowledge research grants funded jointly by Ministero
dell'Istruzione, dell'Universit\`a e della Ricerca (MIUR), by
Universit\`a di Torino and by Istituto Nazionale di Fisica Nucleare
within the {\sl Astroparticle Physics Project} (MIUR contract number: PRIN 2008NR3EBK;
INFN grant code: FA51).


\begin{thebibliography}{99}

\bibitem{Bertone:2010zza}
  G.~Bertone, (Ed.),
  ``Particle Dark Matter: Observations, Models and Searches''.
  

\bibitem{Regis:GC}
  M.~Regis, P.~Ullio,
  Phys.\ Rev.\  {\bf D78 } (2008)  043505.
  [arXiv:0802.0234 [hep-ph]]

\bibitem{GC}
  G.~Bertone, G.~Sigl, J.~Silk,
  Mon.\ Not.\ Roy.\ Astron.\ Soc.\  {\bf 337 } (2002)  98.
  G.~Bertone, M.~Cirelli, A.~Strumia, M.~Taoso,
  JCAP {\bf 0903 } (2009)  009.
  C.~Boehm, J.~Silk, T.~Ensslin,
  [arXiv:1008.5175 [astro-ph.GA]].
  R.~M.~Crocker, N.~F.~Bell, C.~Balazs, D.~I.~Jones,
  Phys.\ Rev.\  {\bf D81 } (2010)  063516.

\bibitem{Haze}
  D.~T.~Cumberbatch et al.,
    [arXiv:0902.0039 [astro-ph.GA]],
  G.~Dobler, D.~P.~Finkbeiner,
  Astrophys.\ J.\  {\bf 680 } (2008)  1222-1234.
  T.~Linden, S.~Profumo, B.~Anderson,
  Phys.\ Rev.\  {\bf D82 } (2010)  063529.
  D.~Hooper, D.~P.~Finkbeiner, G.~Dobler,
  Phys.\ Rev.\  {\bf D76 } (2007)  083012.


\bibitem{Halo}
  E.~Borriello, A.~Cuoco, G.~Miele,
  Phys.\ Rev.\  {\bf D79 } (2009)  023518.
  T.~Delahaye, C.~Boehm, J.~Silk,
  [arXiv:1105.4689 [astro-ph.GA]],
  M.~Regis, P.~Ullio,
  Phys.\ Rev.\  {\bf D80 } (2009)  043525.

\bibitem{Fornengo:2011iq}
  N.~Fornengo, R.~A.~Lineros, M.~Regis, M.~Taoso,
  [arXiv:1110.4337 [astro-ph.GA]].

\bibitem{Fixsen:2009xn}
  D.~J.~Fixsen et al.,
  Astrophys.\ J.\  {\bf 734 } (2011)  5.

\bibitem{Fornengo:2011cn}
  N.~Fornengo, R.~Lineros, M.~Regis and M.~Taoso,
  [arXiv:1108.0569 [hep-ph]].


\bibitem{Zhang:2008rs}
  L.~Zhang, G.~Sigl,
  JCAP {\bf 0809 } (2008)  027.
  Fornengo, Lineros, Regis, Taoso, in preparation.
  

\bibitem{Colafrancesco}
  S.~Colafrancesco, S.~Profumo and P.~Ullio,
  Astron.\ Astrophys.\  {\bf 455} (2006) 21.
  S.~Colafrancesco, S.~Profumo and P.~Ullio,
  Phys.\ Rev.\  D {\bf 75} (2007) 023513.

\bibitem{dwarf}
 M.~Regis et al., in preparation.
\bibitem{atca}
http://www.narrabri.atnf.csiro.au/

\end{thebibliography}
\end{document}